\documentclass[%
article,
twocolumn,
superscriptaddress,
nofootinbib,
amsmath,amssymb,
aps,
]{revtex4-1}

\usepackage{graphicx}
\usepackage{dcolumn}
\usepackage{bm}
\usepackage{hyperref}
\usepackage{float}%

\usepackage{amsmath}
\usepackage{MnSymbol}

\usepackage{makecell}

\begin{document}
	
	
	\title{Discovering Superhard B-N-O Compounds by Iterative Machine Learning and Evolutionary Structure Predictions}
	
	\author{Wei-Chih Chen}
	\email{wcc.weichihchen@gmail.com}
	\affiliation{Department of Physics, University of Alabama at Birmingham, Birmingham, Alabama 35294, USA}
	
	\author{Yogesh K. Vohra}
	\affiliation{Department of Physics, University of Alabama at Birmingham, Birmingham, Alabama 35294, USA}
	
	\author{Cheng-Chien Chen}
	\email{chencc@uab.edu}
	\affiliation{Department of Physics, University of Alabama at Birmingham, Birmingham, Alabama 35294, USA}
		
	\date{\today}
	
	\begin{abstract}
	We search for new superhard B-N-O compounds with an iterative machine learning (ML) procedure, where ML models are trained using sample crystal structures from evolutionary algorithm. We first use cohesive energy to evaluate the thermodynamic stability of varying B$_x$N$_y$O$_z$ compositions, and then gradually focus on compositional regions with high cohesive energy and high hardness. The results converge quickly after a few iterations. Our resulting ML models show that B$_{x+2}$N$_{x}$O$_{3}$ compounds with $x \geq 3$ (like B$_5$N$_3$O$_3$, B$_6$N$_4$O$_3$, etc.) are potentially superhard and thermodynamically favorable. Our meta-GGA density functional theory calculations indicate that these materials are also wide bandgap ($\ge 4.4$ eV) insulators, with the valence band maximum related to the $p$-orbitals of nitrogen atoms near vacant sites. This study demonstrates that an iterative method combining ML and {\it ab initio} simulations provides a powerful tool for discovering novel materials.
	\end{abstract}
	
	\maketitle
	

	
	
	
	
	

\section{Introduction}

With the continuous increase in global demand of superhard materials, searching for new compounds with outstanding hardness and stability is becoming an important research topic~\cite{Zhao_review}. 
Materials of superhardness (with hardness value $H \ge 40$ GPa) can be classified into two major groups.
The first includes transition-metal (TM) ceramics, especially the borides and carbides~\cite{friedrich2011synthesis,akopov2018perspective,burrage2020electronic,burrage2020experimental}.
The second concerns light elements B, C, N, and O~\cite{lundstro1996superhard,kurakevych2009superhard,baker2020first,chakrabarty2020superhard}, which form short and strong covalent bonds.
Various superhard materials can be produced by mixing two or three of these light elements.
Materials in the second group have the advantages of being abundant and low cost, but they may require special synthesis conditions.
Solozhenko {\it et al.} synthesized diamond-like BC$_5$ under high pressure and high temperature (HPHT), and reported a corresponding hardness $H= 71$ GPa~\cite{PhysRevLett.102.015506}. They also found that BC$_5$ has a relatively high fracture toughness and exceptional thermal stability up to 1900 K. Baker {\it et al.} used microwave plasma chemical vapor deposition to synthesize boron-incorporated diamond at low temperature and low pressure. They found a hardness as high as 62 GPa in their cubic phase sample with 7.7 at\% boron content~\cite{baker2018computational}.

Superhard B-C-N compounds also have been studied in the literature.
Cubic BC$_2$N was reported with a superhardness between 62-76 GPa \cite{solozhenko2001mechanical, BC2N_solozhenko2001synthesis, BC2N_zhao2002superhard}, which is harder than cubic boron nitride (c-BN, $H\sim$ 50-70 GPa)~\cite{ZHANG2014607,monteiro2013cubic}.
Experimentally, BC$_2$N remains stable up to 1800 K, which demonstrates its superior thermal stability compared to diamond ($H\sim 100$ GPa). 
Other ternary compounds such as BCN and BC$_9$N in the cubic phase also have been synthesized under extreme conditions~\cite{BCN_liu2011synthesis}.
Recently, Chen {\it et al.} have used machine learning and evolutionary search to discover superhard B-C-N structures. Their newly predicted BC$_{10}$N has an ultra-high hardness $\sim 87$ GPa with a relatively low formation energy~\cite{chen2021machine}.


B-N-O compounds also have attracted considerable attention due to their relevance to several research areas, such as chemical adsorption~\cite{shankar2021gas}, fluorescent dots~\cite{ren2020boron}, water splitting~\cite{xie2012boron}, wide bandgap insulators~\cite{arnold2021composition}, and electrochemical applications~\cite{dussauze2013lithium}. 
Although B-N-O compounds have been synthesized in various forms, including amorphous, porous, layered, thin-film, and nanoparticle,
their mechanical properties remain largely unexplored.
More recently, Bhat {\it et al.} have used hexagonal BN and B$_2$O$_3$ as initial materials to synthesize B$_6$N$_4$O$_3$ under HPHT conditions~\cite{bhat2015high}.
They also conducted computational studies and indicated that the most stable structures contain ordered vacancies in a zinc-blende structure. Such structures have a high bulk modulus of 300 GPa, which implies that B-N-O compounds may be superhard as well~\cite{bhat2015high}.

		\begin{figure}[th!]
	\begin{center}
		\includegraphics[width=\columnwidth]{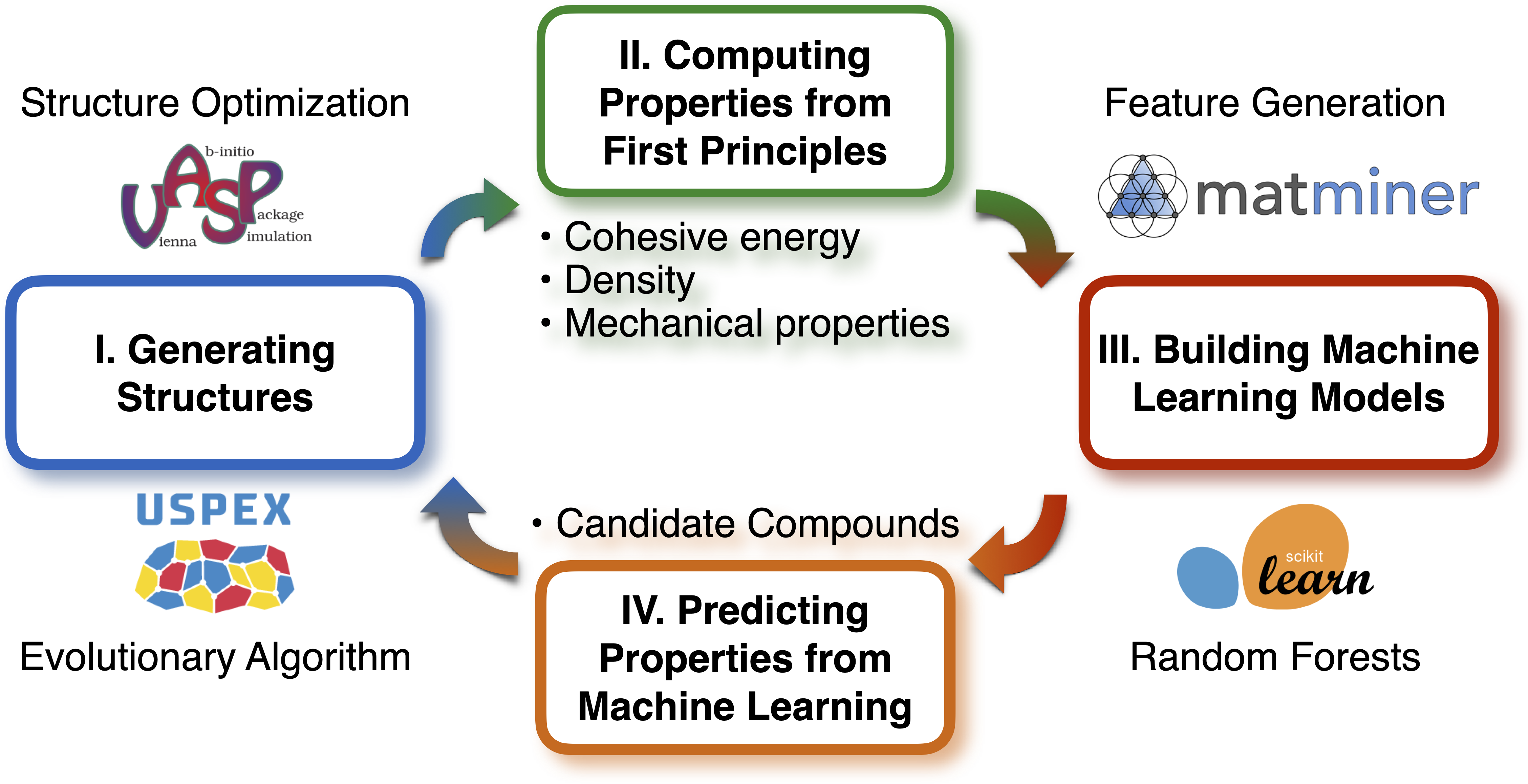}
		\caption{Schematic iterative process of machine learning and structure prediction for superhard B-N-O compounds. {\bf I.} Generating B-N-O crystal structures from evolutionary algorithm implemented in \textsc{USPEX}~\cite{oganov2006crystal, glass2006uspex, lyakhov2013new}. {\bf II.} Computing physical properties from first-principles density functional theory software \textsc{VASP}~\cite{VASP_1,VASP_2}. The target properties include cohesive energy, density, and mechanical properties. {\bf III.} Building machine learning (ML) models with \textsc{scikit-learn}~\cite{scikit-learn}. The ML features from Meredig {\it et al.}~\cite{meredig2014combinatorial} are generated by \textsc{Matminer}~\cite{matminer}. 
		{\bf IV.} Predicting B-N-O physical properties with random forests ML models. Promising compositions with high stability (high cohesive energy) and high hardness are selected for the next round of iterative calculation.
		}
		\label{fig:flowchart}
	\end{center}
	\end{figure}
	
Motivated by the study of Bhat {\it el al.}, here we employ first-principles calculation and machine learning (ML) simulation to search for new superhard B-N-O compounds. Data-driven approaches have proven to be powerful in materials discovery~\cite{kim2021deep, frydrych2021materials,schmidt2019recent}, and several ML models have been applied to find superhard compounds~\cite{chen2021machine,zhang2021finding,mazhnik2020application,avery2019predicting,mansouri2018machine}.
The first important step of building ML models is sample data acquisition.
However, although there are several online computational materials databases~\cite{MP,nomad,OQMD,aflow,JARVIS},
only limited information exists on B-N-O.
The lack of relevant training data may lead to inaccurate prediction of universal ML models, when they are applied to the not-well-explored B-N-O compounds. 
For example, our previous universal ML model~\cite{chen2021machine} would predict a bulk modulus of $\sim 180$ GPa for B$_6$N$_4$O$_3$, which largely deviates from the reported value of $\sim 300$ GPa from first-principles calculation~\cite{bhat2015high}. 
Therefore, we need a different ML scheme for predicting B-N-O systems.

In this paper, we develop an iterative procedure involving crystal structure prediction (CSP), density functional theory (DFT) calculation, and machine learning (ML) simulation, in order to discover new ternary superhard B-N-O materials. The details of CSP, DFT, and ML simulations are discussed in the ``Methods'' section.
Below we first address our iterative calculation process, which involves four major steps as summarized in Fig.  \ref{fig:flowchart}: (i) Generating crystal structures, (ii) Computing physical properties from first principles, (iii) Building ML models, and (iv) Predicting properties using the ML models.

In step (i), we use evolutionary algorithm for CSP to create new B-N-O structures with varying chemical compositions. 
In step (ii), we use DFT to compute the cohesive energy, volumetric density, and elastic tensor, for crystal structures found in the previous step.
The elastic tensor enables the evaluation of a structure's mechanical properties and elastic stability.
The physical properties of stable structures from step (ii) provide the sample data for training ML models.
In step (iii), we build separate ML models respectively for the cohesive energy, density, and hardness.
The cohesive energy helps to determine a structure's thermodynamic stability.
Finally, in step (iv), we construct triangular plots to map out the above properties for arbitrary B-N-O composition, using ML models from the previous step.
Based on the ML predictions, we then select promising chemical compositions with high thermodynamic stability and high hardness as candidates for the next iteration of calculation. We find that our ML models can be gradually improved with this procedure. We stop at the fourth iteration, as the results are quickly converged.
As shown below, using this iterative procedure combining CSP/DFT calculations and ML simulation, we are able to discover several new stable and superhard B-N-O structures.

\begin{figure*}[!th]
	\begin{center}
		\includegraphics[width=1.0\textwidth]{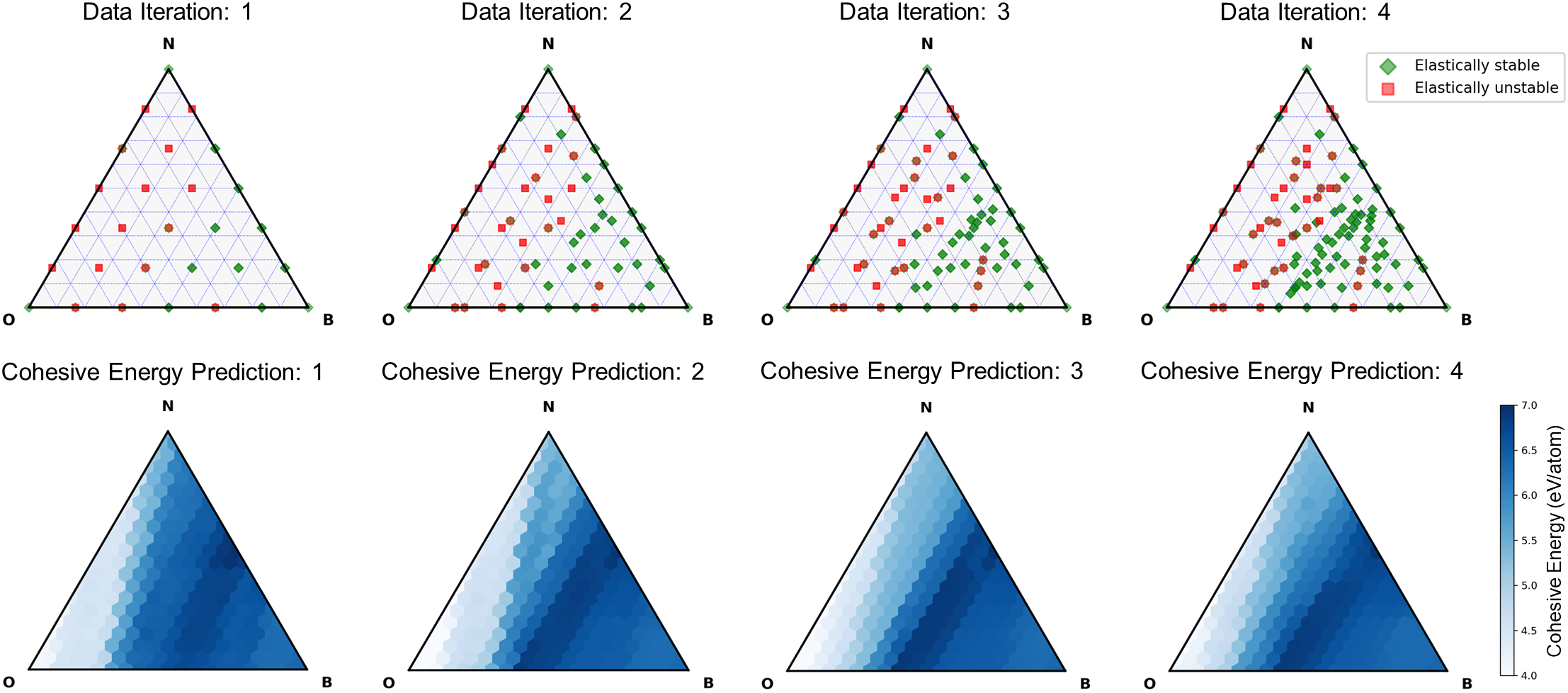}
		\caption{
[Top panels] Distribution of B-N-O compositions in evolutionary structure prediction, and their elastic stabilities computed using density functional theory. Only elastically stable structures (green diamond) are used for constructing machine learning models; elastically unstable structures (red square) are excluded during data selection. 
[Bottom panels] Random forests prediction of cohesive energy. Based on the predicted cohesive energy and hardness (not shown), promising B-N-O compounds are selected for calculation in the next iteration. Ternary graphs are visualized by the \textsc{python-ternary}~\cite{pythonternary} library.
		}
		\label{fig:ml_generations}
	\end{center}
\end{figure*}

\section{Computational Methods}
{\it Crystal structure prediction} (CSP) --
We use CSP to generate B-N-O sample structures, which later serve as the input data for training machine learning (ML) models. The purpose of CSP is to find stable and/or metastable structures of a compound given only its chemical formula~\cite{wang2014perspective,graser2018machine, oganov2019structure}. 
Here, we use the highly powerful and efficient evolutionary algorithm implemented in the USPEX package~\cite{oganov2006crystal, glass2006uspex, lyakhov2013new}. During different generations of evolutionary optimization, new structures are created by heredity (50\%), mutation (30\%), and random (20\%) operators. The enthalpy computed by density functional theory (DFT) is used as the fitness.

We consider varying compositions of ternary B-N-O compounds, with a unit-cell size between 9 to 19 atoms.
We ensure that all of the selected atomic compositions contain an even number of total electrons. This typically leads to a more stable, insulating phase.
Since our goal is to discover superhard compounds, we apply an external pressure $P$ = 15 GPa during the structure searches. Adding a small but finite pressure will help avoid low-hardness layered graphite-like structures.
For each chemical composition, we perform two separate USPEX calculations and search over 1,200 structures. 
The optimized structures from both USPEX searches are further fully relaxed in DFT calculations without any external pressure.
In the end, for a given chemical formula, we have two sample structures for training ML models.

{\it Density functional theory (DFT) calculation} --
We perform DFT calculations using the VASP software~\cite{VASP_1,VASP_2}, which adopts a pseudopotential method and plane-wave basis sets. We use the projector augmented wave (PAW)~\cite{PAW_1,PAW_2} method and generalized gradient approximation (GGA) functional based on the Perdew-Burke-Ernzerhof (PBE) formalism~\cite{PBE}. The kinetic energy cutoff for wavefunction expansion is 520 eV, and the $k$-points are sampled by a $\Gamma$-centered Monkhorst-Pack mesh with a resolution $\sim$ 0.02 $\times$ 2$\pi$/\text{\normalfont\AA}. The convergence criteria of electronic self-consistency and structural relaxation are set to 10$^{-6}$ eV/unit cell and 10$^{-3}$ eV/$\text{\normalfont\AA}$, respectively. 
	
For fully relaxed crystal structures, we further utilize the strain-stress method~\cite{PhysRevB.65.104104} implemented in VASP to compute the elastic constants $C_{ij}$, which in turn help to evaluate the elastic stability and mechanical properties. In particular, the eigenvalues of the elastic tensor (for elastic stability), as well as the Voigt-Reuss-Hill (VRH) averaged ~\cite{voigt1928lehrbuch, reuss1929berechnung, hill1952elastic} bulk modulus $K$ and shear modulus $G$, are computed via the MechElastic python library~\cite{singh2018elastic, singh2021mechelastic}. The Vickers hardness $H$ is also evaluated by using Chen's empirical hardness model~\cite{chen2011modeling}:
	    \begin{equation}
		\label{eq:cohesive}
		\begin{aligned}
			H =  2(k^2G)^{0.585}-3,
		\end{aligned}
	\end{equation}
	where $k =K/G$ is the Pugh's ratio.
Finally, phonon density of states (for dynamical stability) are computed using the \textsc{Phonopy} package~\cite{phonopy}, via the density functional perturbation theory scheme implemented in VASP. Convergence tests of all calculations are examined carefully.
	
{\it Machine learning (ML) prediction --} 
Since the ML predictive capability depends crucially on the training data, we first exclude elastically unstable sample structures, by examining whether the elastic tensor is positive-definite~\cite{BORN-HUANG}. After preparing data from CSP and DFT calculations, we build three ML models respectively for predicting cohesive energy, density, and hardness.

The ML features or descriptors are created using the python library \textsc{Matminer}~\cite{matminer}.
For cohesive energy and density, we adopt the compositional features by Meredig $et$ $al.$~\cite{meredig2014combinatorial}.
A total number of 16 features derived from a given chemical composition are considered: atomic fractions of B, N, and O; mean atomic weight; mean column number; mean row number; range of atomic number; mean atomic number; range of atomic radius; mean atomic radius; range of electronegativity; mean electronegativity; average numbers of $s$ and $p$ electrons; fractions of $s$ and $p$ electrons. For the hardness model, we further consider the volumetric density as an additional feature, in order to improve the model performance.

Our ML models are based on random forests, implemented in the \textsc{scikit-learn} library~\cite{scikit-learn}.
Compared to a single decision tree, ensemble trees can improve model prediction accuracy and meanwhile avoid high variance.
When training the ML models, we use 100 estimators and 10-fold cross validation on 80\% of the samples (the training-validation set).
The maximum tree depth is restricted to 6 layer to further avoid overfitting.

	\begin{figure*}[!th]
	\begin{center}
		\includegraphics[width=\textwidth]{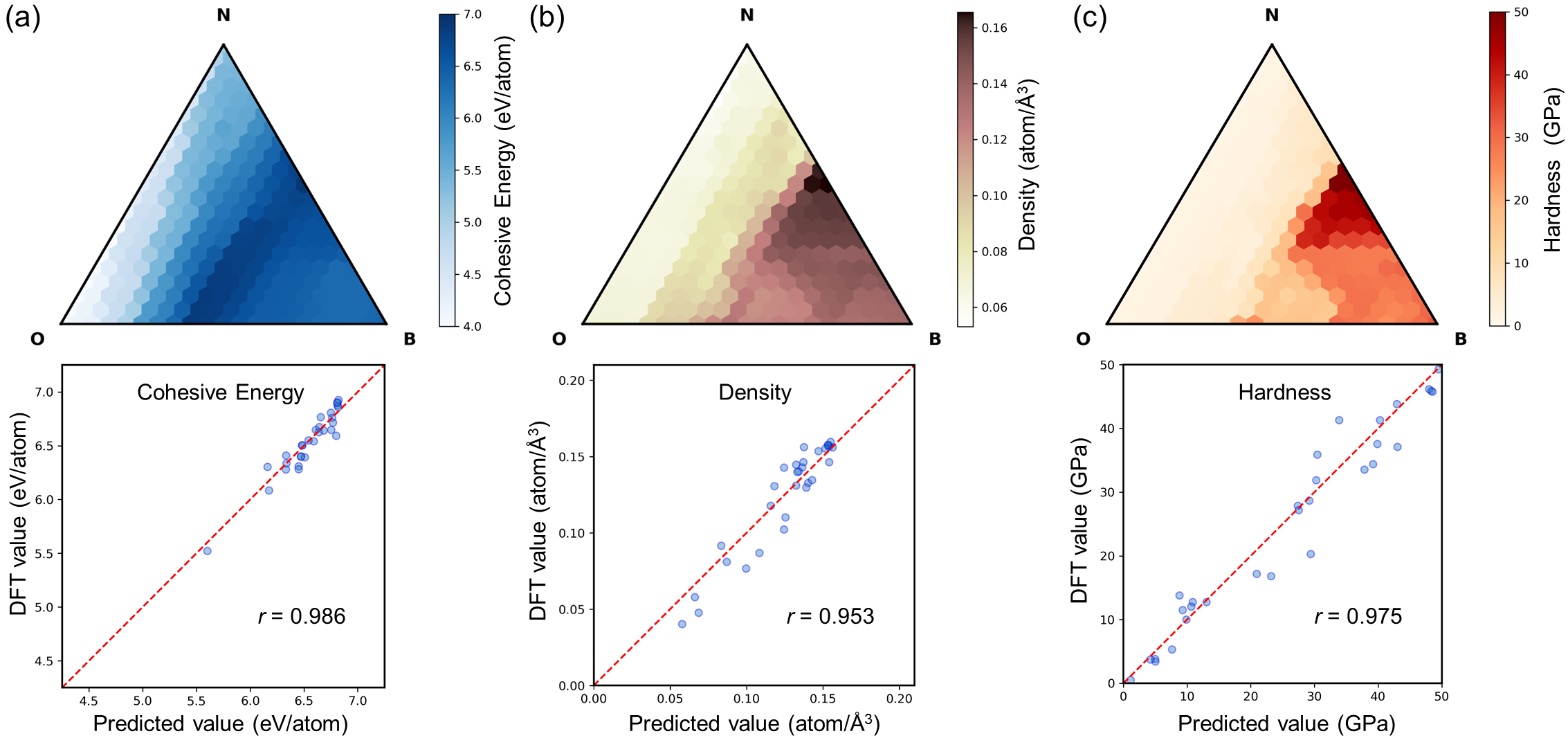}
		\caption{
Prediction [top panels] and evaluation [bottom panels] of random forests machine learning (ML) models for (a) cohesive energy, (b) density, and (c) hardness. The Pearson correlation coefficient ($r$) between the ML predicted value and the density functional theory (DFT) calculation is utilized as the evaluation metric. The ternary graphs indicate that (i) B$_{x+2}$N$_x$O$_3$ [from linear combinations of (BN)$_x$ and B$_2$O$_3$] have higher cohesive energy, and (ii) hardness is strongly correlated with density.
		}
	\label{fig:ml_prediction}
	\end{center}
	\end{figure*}

\section{Results and Discussion}
Figure \ref{fig:ml_generations} shows triangular plots for B-N-O compositions considered in different iterations of our evolutionary structure prediction [top panels], and the cohesive energy predicted by our machine learning (ML) random forests models [bottom panels]. In the first iteration, we uniformly sample the B-N-O compositional space, with 12-atom unit cells for binary and ternary compounds. Corner points of the graphs correspond to elemental compounds. For example, pure B here represents $\alpha$-B$_{12}$.

In the first iteration of data sampling, the ratio between elastically stable and unstable structures is roughly 1:1. Even though the sampling grid is coarse, it can be seen that ternary compounds composed of mostly nitrogen and oxygen tend to be elastically unstable. In contrast, the systems with high boron content are elastically more stable. To ensure high prediction accuracy, we only consider elastically stable structures to build our ML models. In the four iterations, the numbers of elastically stable structures involved in building ML models are 27, 78, 104, and 153, respectively.

The ML prediction of cohesive energy in Fig. \ref{fig:ml_generations} also indicates that compounds with more boron atoms show higher cohesive energy (i.e. higher thermodynamic stability). The behaviors of elastic stability and thermodynamic stability are thereby consistent with each other. Therefore, starting from the second iteration, we do not consider compositions with simultaneous high nitrogen and high oxygen contents. Instead, we manually select B-N-O compositions based on ML prediction from the previous iteration, by gradually zooming in compositional space with both high predicted cohesive energy and high hardness. With increasing number of iterations, the ML results then have higher and higher accuracy and resolution. Importantly, we find that a region connecting BN and B$_2$O$_3$ (forming B$_{x+2}$N$_x$O$_3$) are thermodynamically favorable with high cohesive energy. This result is consistent with the recently reported B$_6$N$_4$O$_3$ compound by Bhat {\it et al.}, where a mixture of hexagonal BN and B$_2$O$_3$ is used as starting materials for HPHT synthesis~\cite{bhat2015high}.

Figure \ref{fig:ml_prediction} shows the predictions of our random forests ML models respectively for cohesive energy, density, and hardness, built from training samples in the fourth iteration of evolutionary structure searches. Here, 80\% of the samples are used as the training-validation set to construct random forests models, and the remaining 20\% are used as the test set for a final unbiased evaluation of model performance. Our ML model for the cohesive energy shows high prediction accuracy, with a Pearson correlation coefficient $r = 0.986$ between the ML values and actual DFT calculations. The model performances for predicting density and hardness are also comparable, with $r$ scores of 0.953 and 0.975, respectively.

	\begingroup
	\setlength{\tabcolsep}{6pt} 
	\renewcommand{\arraystretch}{1.5} 
	\begin{table*}[t]
		\caption{Physical properties of superhard B-N-O compounds with cohesive energy $>$ 6.75 eV discovered in this study: Density $\rho$ (atom/$\text{\normalfont\AA}^3$), bulk modulus $K$ (GPa), shear modulus $G$ (GPa), Young's modulus $E$ (GPa), Pugh's ratio $k=K/G$, Poisson's ratio $\nu$, universal elastic anisotropy $A^U$, hardness $H$ (GPa), bandgap $E_g$ (eV), cohesive energy $E_{coh}$ (eV/atom), and formation energy $E_{form}$ (meV/atom). The bandgaps are computed respectively with the standard Perdew-Burke-Ernzerhof (PBE)~\cite{PBE} functional and the Tran-Blaha modified Becke-Johnson (TB-mBJ)~\cite{TB,mBJ} exchange potential for improved bandgap estimation.}
		\label{tab:table1}
		\begin{tabular}{l|c|c|c|c|c|c|c|c|c|c|r} 
			Crystal   & $\rho$ & $K$ & $G$ & $E$ & $k$ & $\nu$ & $A^U$ & $H$ ($H^{exp}$) & $E_g^{PBE}$/$E_g^{mBJ}$ ($E_g^{exp}$) & $ E_{coh}$ & $E_{form}$ \\
			\hline
			B$_5$N$_3$O$_2$      & 0.151 & 286  & 257 & 593 & 1.115 & 0.155 & 0.294 & 42 & 4.3 / 5.5 & 6.761 & \\
			B$_5$N$_3$O$_3$ (WZ) & 0.150 & 279  & 255 & 586 & 1.097 & 0.150 & 0.088 & 43 & 6.3 / 7.6 & 6.935 & 83 \\
			B$_6$N$_4$O$_2$      & 0.156 & 322  & 284 & 658 & 1.136 & 0.160 & 0.121 & 44 & 3.0 / 4.4 & 6.810 & \\
			B$_6$N$_4$O$_3$ (ZB) & 0.155 & 292  & 268 & 616 & 1.091 & 0.149 & 0.311 & 45 & 4.5 / 5.7 & 6.900 & 120 \\
			B$_7$N$_5$O$_2$      & 0.146 & 272  & 241 & 559 & 1.126 & 0.157 & 0.336 & 40 & 3.6 / 4.6 & 6.776 & \\
			B$_7$N$_5$O$_3$ (ZB) & 0.158 & 306  & 283 & 649 & 1.081 & 0.147 & 0.276 & 47 & 4.1 / 5.3 & 6.904 & 117 \\
			B$_9$N$_7$O$_2$      & 0.161 & 330  & 296 & 683 & 1.114 & 0.154 & 0.139 & 46 & 3.2 / 4.5 & 6.802 & \\
			B$_9$N$_7$O$_3$ (ZB) & 0.160 & 318  & 301 & 688 & 1.055 & 0.140 & 0.245 & 50 & 3.9 / 5.1 & 6.926 & 97 \\
			cubic BN             & 0.168 & 373 & 383 & 856 & 0.976 & 0.118 & 0.172 & 64 (50$-$70)\cite{ZHANG2014607} & 4.5 / 5.3 (6.36)\cite{Evans_2008}  & 7.028 & 0 \\
			B$_2$O$_3$ (P3$_1$21)& 0.101 & 35 & 33   & 75  & 1.049 & 0.127 & 2.347 & 12 (1.5)\cite{Mukhanov2008} & 6.3 / 8.9 ($>$10)\cite{Sayyed2018} & 7.008 & 0
		\end{tabular}
	\end{table*}
	\endgroup

In general, a material's hardness is correlated with its volumetric density~\cite{de2015charting}. For example, cubic diamond as the hardest material also has the largest volumetric density at ambient condition. Therefore, to enhance the ML prediction accuracy, we also consider density as an additional feature when building the hardness model. We first use only compositional features to train a density ML model, which in turn is utilized to generate density feature for the subsequent hardness ML model. As shown in Figs. \ref{fig:ml_prediction}(b) and \ref{fig:ml_prediction}(c), the ML predicted density and hardness values indeed show nearly a linear relationship. 

Our iterative calculation combining evolutionary structure search and ML prediction show that B-N-O compounds can exhibit superhardness in a compositional region around BN. With the aid of the ternary graphs shown in Fig. \ref{fig:ml_prediction}, we can directly inspect the sample crystal structures generated from the evolutionary algorithm. Table \ref{tab:table1} lists various superhard B-N-O compounds discovered in our study. In particular, we find mainly two kinds of ternary compositions. In the first kind, the number of boron is equal to the total number of nitrogen and oxygen (e.g. B$_5$N$_3$O$_2$). The second kind contains one more oxygen, where the number of boron is equal to the total number of nitrogen and oxygen minus 1 (e.g. B$_5$N$_3$O$_3$). 

Using DFT calculations, we have studied various physical properties of B-N-O compounds listed in Table \ref{tab:table1}. In particular, their volumetric densities $\rho$ (atom/$\text{\normalfont\AA}^3$) are overall positively correlated with the hardness values, computed by Chen's empirical hardness model based on the bulk modulus $K$ and shear modulus $G$~\cite{chen2011modeling}.
Moreover, by examining the electronic density of states (DOS), we find these new superhard B-N-O materials are insulators, with wide bandgaps $\ge$ 3.0 eV based on the PBE functional. It is known that GGA functionals like PBE will tend to underestimate the bandgaps. For a more accurate estimation, we further consider the Tran-Blaha modified Becke Johnson (TB-mBJ)~\cite{mBJ_1,mBJ_2} meta-GGA exchange potential. Our TB-mBJ results show that the superhard B-N-O compounds under study are wide-bandgap insulators with a gap size $\ge$ 4.4 eV.

	\begin{figure*}[!th]
	\begin{center}
		\includegraphics[width=\textwidth]{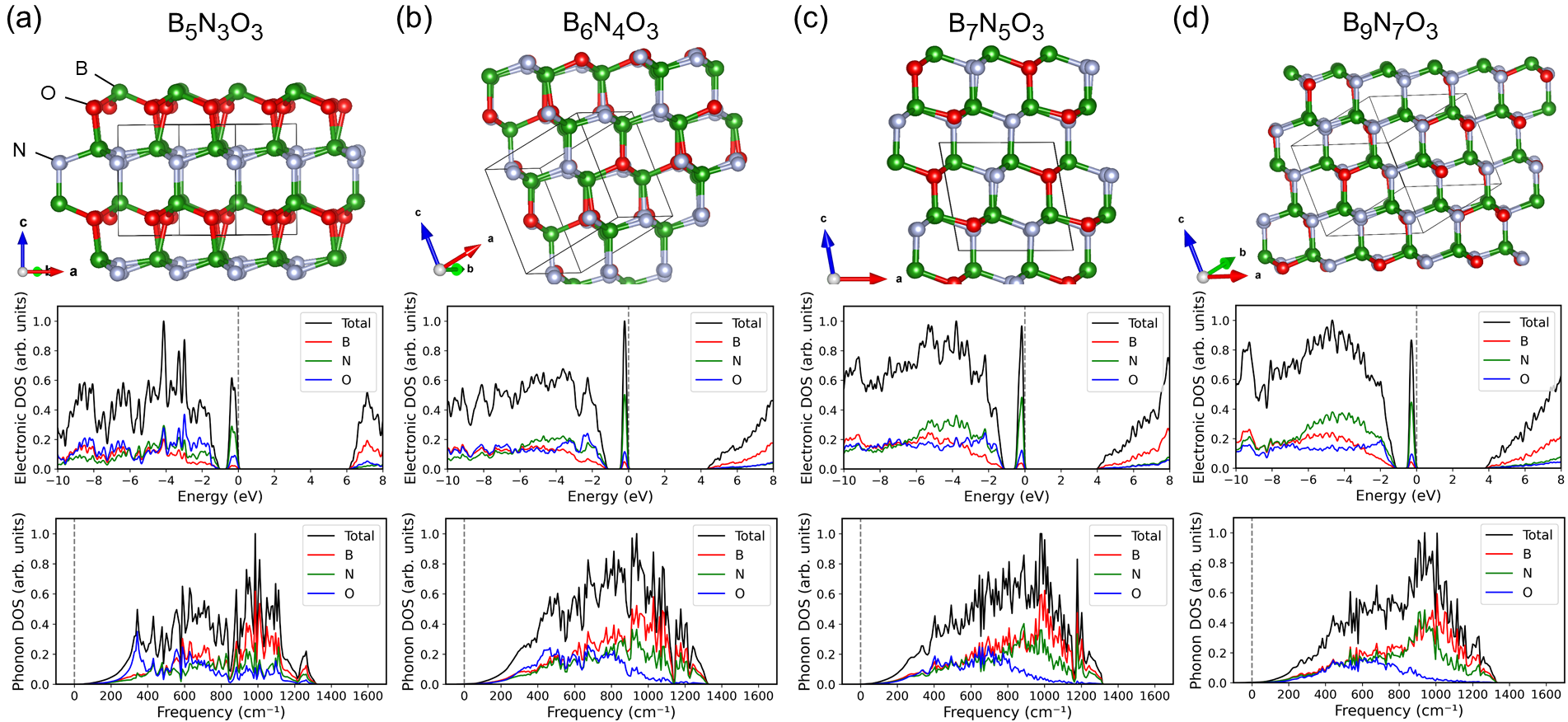}
		\caption{\label{fig:dft}
Predicted crystal structures [top panels], electronic density of states (DOS) [middle panels], and phonon DOS [bottom panels]  for (a) B$_5$N$_3$O$_3$, (b) B$_6$N$_4$O$_3$, (c) B$_7$N$_5$O$_3$, and (d) B$_9$N$_7$O$_3$. The results are obtained using the Perdew-Burke-Ernzerhof (PBE) functional. All four compounds are wide-bandgap insulators, with a peak at the valence band maximum originating from $p$-orbitals of nitrogen atoms near the vacant sites (see Fig. \ref{fig:fig5}). The phonon spectra show only positive modes, indicating dynamical stability of all four compounds. The crystal structures are visualized by the VESTA software~\cite{momma2011vesta}.
		}
	\end{center}
	\end{figure*}

We next evaluate the thermodynamic stabilities of the newly predicted superhard B-N-O compounds, by comparing their cohesive energy:
    \begin{equation}
	\label{eq:cohesive}
	\begin{aligned}
		E_{coh} =  \frac{xE(B_{atom}) + yE(N_{atom}) + zE(O_{atom}) - E(B_xN_yO_z)} {x+y+z}.
	\end{aligned}
	\end{equation}
With the total energies of isolated atoms as references, $E_{coh}$ can be regarded as the formation energy with a minus sign. Therefore, the higher the cohesive energy, the higher the thermodynamic stability of a material. The second kind of compounds B$_{x+2}$N$_x$O$_{3}$ are found to exhibit a higher cohesive energy than the first kind. Moreover, the cohesive energies of c-BN and B$_2$O$_3$ are even higher than the predicted B-N-O compounds, which are thereby metastable at ambient condition, but may be stabilized under HPHT synthesis conditions~\cite{bhat2015high}.

For B$_{x+2}$N$_x$O$_{3}$, we also calculate the formation energy using the total energies of c-BN and B$_2$O$_3$ as references. In particular, the formation energy for these B-N-O compounds can be computed by
	\begin{equation}
	\label{eq:formation}
	\begin{aligned}
		E_{form} =  \frac{E(B_{x+2}N_xO_{3}) - xE(c-BN) - E(B_2O_3) } {2x+5}.
	\end{aligned}
	\end{equation}
Based on this formula, B$_5$N$_3$O$_3$ exhibits the lowest formation energy among the B-N-O compounds under study. A corresponding superlattice structure computed by evolutionary algorithm for B$_5$N$_3$O$_3$ is shown in Fig. \ref{fig:dft}(a). The result is consistent with the report from Bhat {\it et al.} that models of ordered structure would agree better with their experiments~\cite{bhat2015high}.

\begin{figure}[!th]
	\begin{center}
		\includegraphics[width=1.0\columnwidth]{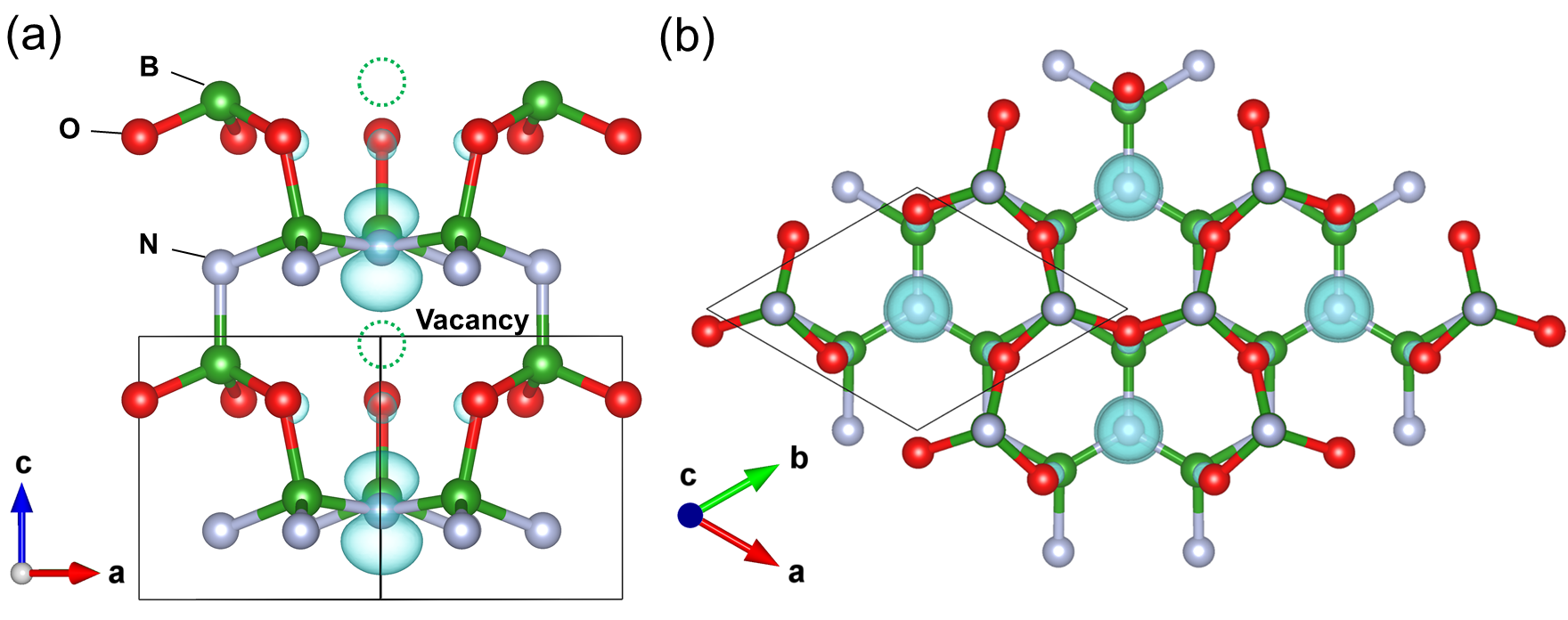}
		\caption{B$_5$N$_3$O$_3$ crystal structure: (a) side view, and (b) top view. The unit cell is indicated by thin solid lines. The vacant sites around three oxygen atoms and one nitrogen atom are emphasized by dotted circles. The electron charge contour (cyan color) corresponds to the valence band maximum in the electronic density of states, and it is mainly contributed by nitrogen $p$-orbitals near the vacant sites.
		}
		\label{fig:fig5}
	\end{center}
	\end{figure}

Based on Table \ref{tab:table1}, we select a few more stable B$_{x+2}$N$_x$O$_{3}$ compounds to further characterize their properties. Figure \ref{fig:dft} shows the crystal structures, electronic density of states (DOS), and phonon DOS, respectively for B$_5$N$_3$O$_3$, B$_6$N$_4$O$_3$, B$_7$N$_5$O$_3$, and B$_9$N$_7$O$_3$. Among these compounds, only B$_5$N$_3$O$_3$ assumes a wurtzite structure, and all the others have a zinc-blende structure. Compared to BN, B$_{x+2}$N$_x$O$_{3}$ contains one less boron atom (due to B$_2$O$_3$), leading to vacant boron sites as indicated by the cyan circles in Fig. \ref{fig:fig5}. 

As seen in the middle panels of Fig. \ref{fig:dft}, the three zinc-blende structures have similar electronic profiles around the Fermi level, with a bandgap $\gtrsim$ 4 eV. The electronic DOS of the wurtzite B$_5$N$_3$O$_3$ is also similar, but it exhibits an even larger bandgap $\ge$ 6 eV. This result is unusual, because in binary BN structures, wutzite BN has a smaller bandgap than that of cubic BN, which is distinct from the ternary B-N-O compounds studied here. We also find that in the zinc-blende structures, the bandgap increase with increasing oxygen content. This result is reasonable, as B$_2$O$_3$ has an extremely large band gap $\ge$ 10 eV~\cite{B2O3_10eV}, compared to that of 6 eV in c-BN. Overall, introducing oxygen atoms into binary BN compounds can substantially alter the electronic structures, which also could cause changes in thermal and chemical stabilities. Finally, the phonon DOS (bottom panels of Fig. \ref{fig:dft}) show only positive phonon modes, ensuring that these B-N-O compounds are dynamically stable.

Finally, we note that the electronic DOS for all compounds in Fig. \ref{fig:dft} show a sharp peak at the valence band maximum.
By studying the local electronic DOS, we find the sharp peak is mainly contributed by localized $p$-orbitals of nitrogen atoms next to the vacant boron sites. The electron charge distribution corresponding to the sharp peak in B$_5$N$_3$O$_3$ is shown as cyan colors in Fig. \ref{fig:fig5}.	

\section{Conclusion}
We have developed an iterative machine learning procedure to discover new superhard B-N-O compounds, where the input training samples are generated from evolutionary structure searches and density functional theory calculations.
Our combined machine learning and first-principles results revealed several stable and superhard B-N-O compounds of chemical compositions B$_{x+2}$N$_{x}$O$_3$ ($x \ge$ 3), with hardness values $\gtrsim$ 45 GPa. 
We also found these newly predicted B-N-O systems are all wide bandgap insulators, with gap size $\ge$ 4 eV based on the Perdew-Burke-Ernzerhof GGA functional. Our additional meta-GGA calculations indicate that their actual bandgaps could be even larger. The electronic density of states for B$_{x+2}$N$_{x}$O$_3$ all show a prominent peak around the valence band maximum, and it is related to localized $p$-orbitals of nitrogen atoms near vacant boron sites. 
Since B$_5$N$_3$O$_3$ has already been reported in the literature, other B$_{x+2}$N$_{x}$O$_3$ compounds in principle can be synthesized by high-pressure high-temperature techniques or by chemical vapor deposition methods. We expect these newly-discovered superhard B-N-O materials have a wide range of applications in extreme environments, and they may outperform diamond or cubic boron nitride in oxidizing environment or in humid condition at high temperature.

\section*{Acknowledgments}
This research is supported by the U.S. National Science Foundation (NSF) under award OIA-1655280.
The calculations were performed on the Frontera computing system at the Texas Advanced Computing Center. Frontera is made possible by NSF award OAC-1818253.\\

\section*{Author Contributions}
W.-C.C. performed machine learning simulations and analyzed the results. W.-C.C. performed crystal structure prediction and density functional theory calculations. Y.K.V. and C.-C.C. conceived and supervised the project. W.-C.C. and C.-C.C. wrote the manuscript. All authors helped revise the manuscript.\\

\section*{Data and code availability}
Python codes and data needed for reproducing our machine learning results are downloadable at https://github.com/weichihuab/ML\_B-N-O.
Crystal structure information from evolutionary prediction and density functional theory calculation is also available in the above hyperlink.

\section*{Competing Interests}
The authors declare no competing interests.

\bibliography{bibfile_BNO}
	
\end{document}